\documentclass[aps,pre,preprint,superscriptaddress,showpacs]{revtex4-1}

\usepackage{amssymb}
\usepackage{color}
\usepackage{graphicx}
\usepackage{amsmath}
\usepackage{array}
\usepackage{mathrsfs}
\usepackage{times}
\usepackage{subeqnarray}
\usepackage{cases}
\usepackage{bm}
\usepackage{float}

\begin{document}

\title{Huygens' synchronization experiment revisited: Luck or skill?}

\author{Jiao Yang}
\affiliation{School of Physics and Information Technology, Shaanxi Normal University, Xi'an 710062, China}
\author{Yan Wang}
\affiliation{School of Physics and Information Technology, Shaanxi Normal University, Xi'an 710062, China}
\author{Yizhen Yu}
\email[Email address: ]{yzyu@snnu.edu.cn}
\affiliation{School of Physics and Information Technology, Shaanxi Normal University, Xi'an 710062, China}
\affiliation{National Demonstration Center for Experimental X-physics Education, Shaanxi Normal University, Xi'an 710062, China}
\author{Jinghua Xiao}
\affiliation{School of Science, Beijing University of Posts and Telecommunications, Beijing 100876, China}
\author{Xingang Wang}
\email[Email address: ]{wangxg@snnu.edu.cn}
\affiliation{School of Physics and Information Technology, Shaanxi Normal University, Xi'an 710062, China}
\affiliation{National Demonstration Center for Experimental X-physics Education, Shaanxi Normal University, Xi'an 710062, China}

\begin{abstract}
353 years ago, in a letter to the Royal Society of London, Christiaan Huygens described ``an odd kind of sympathy" between two pendulums mounted side by side on a wooden beam, which inspired the modern studies of synchronization in coupled nonlinear oscillators. Despite the blooming of synchronization study in a variety of disciplines, the original phenomenon described by Huygens remains a puzzle to researchers. Here, by placing two mechanical metronomes on top of a freely moving plastic board, we revisit the synchronization experiment conducted by Huygens. Experimental results show that by introducing a small mismatch to the natural frequencies of the metronomes, the probability for generating the anti-phase synchronization (APS) state, i.e., the ``odd sympathy" described by Huygens, can be clearly increased. By numerical simulations of the system dynamics, we conduct a detailed analysis on the influence of frequency mismatch on APS. It is found that as the frequency mismatch increases from $0$, the attracting basin of APS is gradually enlarged and, in the meantime, the basin of in-phase synchronization (IPS) is reduced. However, as the frequency mismatch exceeds some critical value, both the basins of APS and IPS are suddenly disappeared, resulting in the desynchronization states. The impacts of friction coefficient and synchronization precision on APS are also studied, and it is found that with the increases of the friction coefficient and the precision requirement of APS, the critical frequency mismatch for desynchronization will be decreased. Our study indicates that, instead of luck, Huygens might have introduced, deliberately and elaborately, a small frequency mismatch to the pendulums in his experiment for generating the ``odd sympathy". 

\end{abstract}
\date{\today }
\maketitle

\section{INTRODUCTION}\label{section1}

As a universal concept in nonlinear science, synchronization has been extensively studied by researchers in different fields~\cite{Kuramoto,PRK:2001,Strogatz:2003}. Roughly, synchronization refers to as the coherent motion between coupled nonlinear oscillators, which occurs only when the coupling strength between the oscillators is larger to some critical value. Depending on the specific form of the coherence, synchronization can be classified into different types~\cite{PRK:2001}, including complete synchronization, phase synchronization, generalized synchronization, etc. In complete synchronization, the trajectories of the oscillators are converged into a single one in the phase space, and the states of the oscillators are identical at any instant of the system evolution. In phase synchronization, only the phases of the oscillators are entrained, while the amplitudes of the oscillators remain uncorrelated; and in generalized synchronization, the states of the oscillators are constrained by a function, although the function might not be explicitly given. Among different types of synchronization, phase synchronization is special in that it describes the coherent motion in parallel to the trajectories (which, in many realistic situations, is more concerned than complete and generalized synchronization) and, more importantly, can be realized in a variety of systems, e.g., systems made up of non-identical or different types of oscillators~\cite{Kuramoto,PRK:2001,Strogatz:2003}. For all types of synchronization, a general finding in previous studies is that by increasing the mismatch of the oscillator parameters, the propensity for synchronization will be deteriorated~\cite{SYN:Rev,Kuramoto:Rev,SJ:EPL,AS:EPL}.

The study of synchronization can be traced back to the discovery of the Dutch scientist Christiaan Huygens in the $17$th century. In his letter to the Royal Society of London in February of $1665$~\cite{Huygens}, Huygens described that two mechanical clocks hanging from a common beam always end up with  ``an odd kind of sympathy" in which the clocks are swinging in exactly the same frequency but are $180^{\circ}$ out of phase. Although Huygens' experiment failed in solving the longitude problem for navigation, the intriguing phenomenon he discovered, now known as anti-phase synchronization (APS), has inspired the study of synchronization of coupled nonlinear oscillators in the past decades~\cite{Kuramoto,PRK:2001,Strogatz:2003}. Even now, the experiment designed by Huygens, or its variants, is still employed as an economic-yet-effective setup for exploring the collective dynamics of coupled oscillators, as well as for classroom demonstration of synchronization~\cite{2002 bennett,2002 Pantaleone,AYK:2003,2009 Ulrichs,ClockSyn:nonidentical,2012 Kapitaniak,2012 Wu,2013 Hu,2013 Martens, 2013 Boda,PendulaSyn:EPJ,2014 Kapitaniak,2014 Wu,OHM:2015,2015 Jia,2015 Song,WAR:2017,ZJ:2017,2009 Ulrichs}. Besides experiments, studies have been also carried out in recent years on the necessary conditions for generating APS in coupled pendulums~\cite{2002 bennett,AYK:2003,ClockSyn:nonidentical,2002 Pantaleone,2009 Ulrichs,2012 Wu,2014 Wu,2009 Ulrichs,2012 Kapitaniak}. With high-precision modern equipments, Bennett \emph{et.al.} re-examined Huygens' experiment and found that, to generate APS, the frequencies of the pendulums should be very close, with a precision that can not be achieved in the $17$th century~\cite{2002 bennett}. It is thus suggested that Huygens' observation of ``odd sympathy" depended somewhat on luck. Kanunnikov and Lamper studied the impact of nonlinear interaction between the beam and pendulums on APS, and found that the pendulums can not be exactly out of phase by $180^{\circ}$~\cite{AYK:2003}. The similar phenomenon has been also observed by Czolczynski \emph{et.al.} in coupled pendulums of different masses~\cite{ClockSyn:nonidentical}. A variant of Huygens' experiment is proposed by Pantaleone, in which two metronomes are placed on a freely moving base~\cite{2002 Pantaleone}. It is found that the metronomes are typically developed to in-phase synchronization (IPS), with APS only observable under special conditions. By the theoretical model proposed by Pantaleone, Ulrichs \emph{et al.} studied numerically the synchronization of coupled metronomes, and pointed out that APS is always unstable and existing only in the transient processes~\cite{2009 Ulrichs}. Wu \emph{et al.} studied the attracting basin of APS in the phase space and found that, by adjusting slightly the friction between the base and its support, the basin can be significantly enlarged, making APS observable for the general initial conditions~\cite{2012 Wu,2014 Wu}. Despite the extensive studies on pendulum synchronization, the ``odd sympathy" discovered by Huygens remains as a puzzle to researchers: Did Huygens generate APS by luck? 

\begin{figure}
\begin{center}
\includegraphics[width=0.6\linewidth]{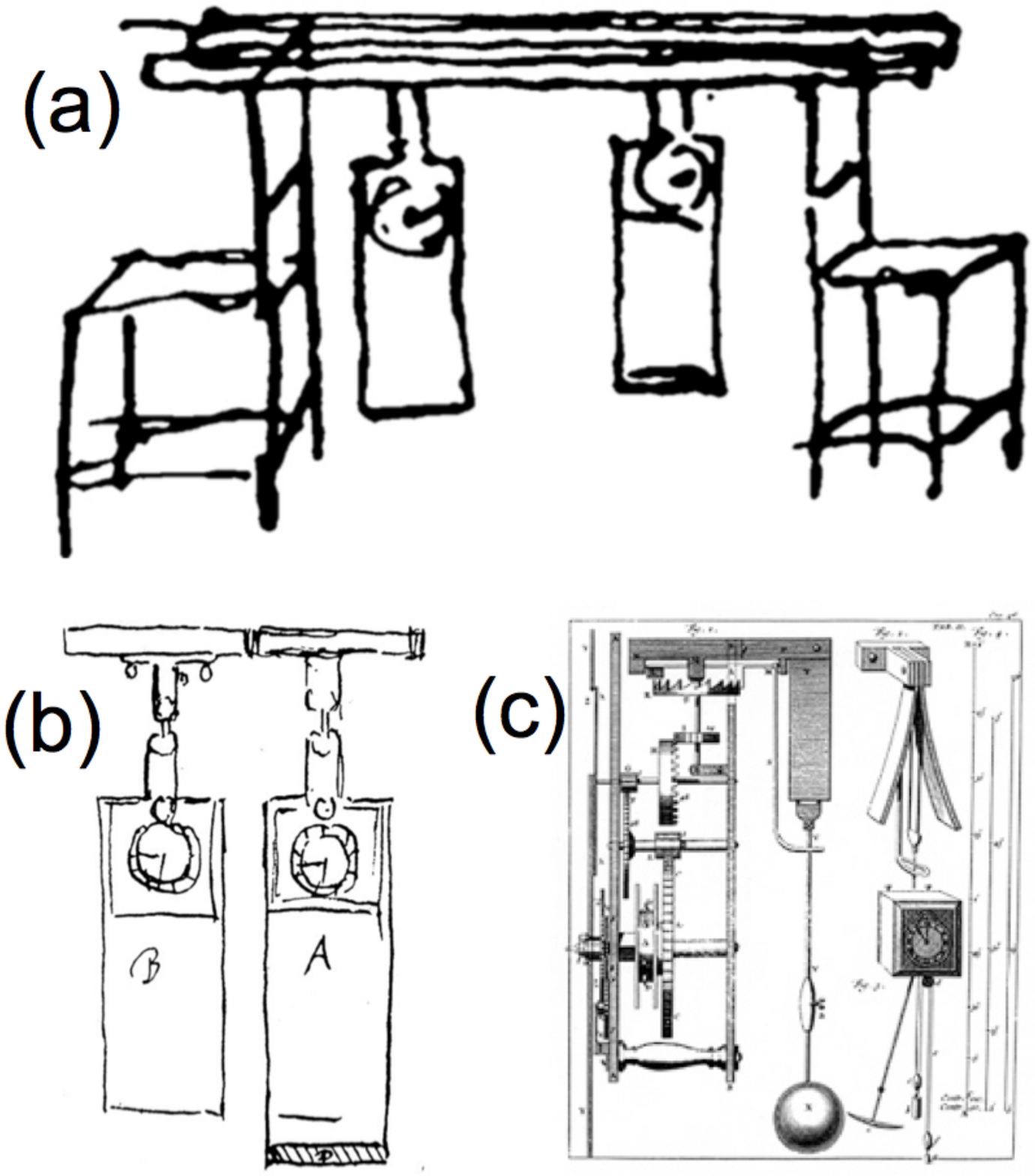}
\caption{(a) Sketch of Huygens' experiment that generates APS. (b) Sketch of the pendulums used in Huygens' experiment. The two pendulums are apparently different from each other. (c) Sketch of the verge pendulum clock. All sketches were drawn by Huygens.} \label{fig1}
\end{center}
\end{figure}

As one of the greatest scientists in the $17$th century, Huygens is famous for his rigorous design of various instruments and devices. The designs are usually accompanied with hand-drawing sketches, which, in most cases, are elaborate and accurate. In describing the experiment that generates the ``odd sympathy", Huygens drew also a sketch in his laboratory notebook, as shown in figure \ref{fig1}(a)~\cite{Huygens}. To give more details about the experiment, Huygens drew separately another sketch for the pendulums [figure \ref{fig1}(b)]. Apparently, the two pendulums are of different lengths. In particular, a weight is attached at the bottom of pendulum A (to stabilize the case), while this weight is absent for pendulum B. This ``careless" drawing is contrary to the way Huygens usually behaves, say, for example, the sketch of the verge pendulum clock drawn in his book \emph{Horologium Oscillatorium} [figure \ref{fig1}(c)]. By the time Huygens did the synchronization experiment, he have already derived the formula for the period of physical pendulum, so it is hard to believe that the difference of the pendulums is made by Huygens casually. Intrigued by this serendipity, in the present work we revisit Huygens' experiment by introducing a small mismatch to the natural frequencies of the pendulums, and study the impact of the frequency mismatch on APS. We are able to demonstrate experimentally and argue theoretically that below a critical frequency mismatch, the attracting basin of APS in the phase space is gradually enlarged as the frequency mismatch increases. Our study therefore suggests that the masses of the pendulums might be deliberately set as different by Huygens in order of generating APS.         

\section{Experimental study}

\begin{figure}
\begin{center}
\includegraphics[width=0.65\linewidth]{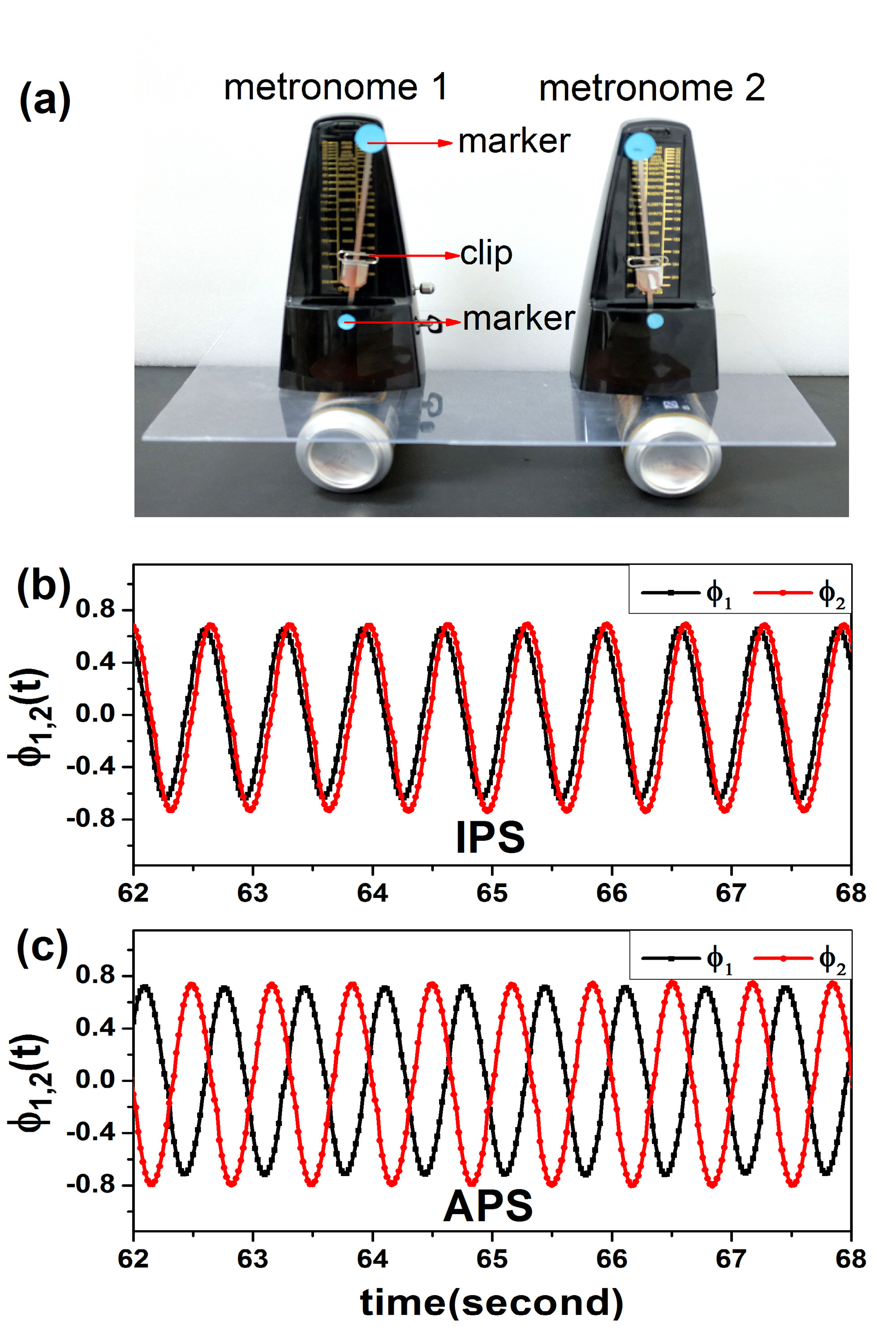}
\caption{(Color online) (a) The experimental setup. (b) The in-phase synchronization state. (c) The anti-phase synchronization state. Experiments are recorded by a camera and analyzed by the software MATLAB.} \label{fig2}
\end{center}
\end{figure}

Our experimental setup is presented in figure \ref{fig2}(a). Following the design of Pantaleone \cite{2002 Pantaleone}, we place two mechanical metronomes on top of a freely moving plastic board. The plastic board is supported by two empty soda cans. The metronomes (Cherub WSM-330) are of nearly identical parameters (parameter mismatch less than 1\%), with the energy supplied by a hand-wound spring. The frequency of the metronome can be adjusted by changing the position of the mass on the pendulum bob, ranging from 40 ticks per minute (largo) to 208 ticks per minute (prestissimo) in a discrete fashion. In the meanwhile, a paper clip is attached to each pendulum bob, which will be used to adjust the metronome frequency slightly. Two small paper markers are sticked on each metronome, with one sticked on the top of the pendulum bob, and the other one sticked at the pivot center. The whole setup is then placed on top of the desk which is horizontal to the ground. (As the plastic board is rigid and supported by identical cans, it is also horizontal to the ground). This experimental setup of coupled metronomes is easy to assemble and can generate robust IPS and APS within a few minutes, making it an excellent experiment for classroom demonstration.

We wish to highlight the function of the paper clips in our experiment. The clips are of the same length ($29$ mm $\pm 1\%$) and weight ($0.5$ g $\pm 1\%$),  and are placed just above the masses on the pendulum bobs. Comparing to the masses (which are about $20$ g), the clips are much light and, by moving them along the bob, provide a slight change to the natural frequency of the metronomes. Specifically, as the clip lifts up, the frequency of the metronome will be gradually decreased, with a precision difficult to be achieved by moving the mass. In our experiment, we fix the position of the clip on metronome $1$, while lifting the clip on metronome 2 to enlarge the frequency mismatch between the metronomes. Throughout our experimental study, we fix the positions of the masses at presto for both metronomes (184 ticks per minute).  

The motions of the pendulums are recorded by the camera of a smart cell phone ($50$ frames per second), and the video is analyzed by the software MATLAB. In analyzing the video, the markers are used as the reference points for acquiring the instant phases, $\phi_{1,2}(t)$, of the metronomes (see Appendix for a detailed description on the acquisition of instant phases). In doing the experiment, we shift the bobs to random (different) initial phases and then release them from the static states. For each experiment, the video is started after a transient period of $60$ seconds, and is lasted for $180$ seconds. If during the recorded period the phase difference between the metronomes, $\delta \phi=|\phi_1-\phi_2|$, is always smaller to a predefined threshold $\Delta \phi=0.3$ rad, the system is judged as reaching IPS; if $|\delta \phi-\pi|<\Delta \phi$, we regard the system as reaching APS; otherwise, the system is regarded as in the non-synchronization state. 

We start by placing the clips at the same position (just above the masses), and measure the natural frequencies, $f_{1,2}$, of the isolated metronomes by putting them on the desk. The results are averaged over $3$ realizations. Data analysis shows that $f_1\approx 1.493$ Hz and $f_2\approx 1.496$ Hz. That is, the frequency mismatch between the metronomes is $\Delta f\equiv f_2-f_1\approx 3.0\times 10^{-3}$ Hz. We then put the metronomes on the plastic board, and repeat the same experiment for $100$ realizations, with the initial phases of the metronomes being randomly chosen [within the range $(-0.5,0.5)$] in each realization. The results are summarized in Tab. \ref{table1}, which shows that the system reaches IPS and APS by the probabilities of $51\%$ and $49\%$, respectively, and no non-synchronization state is observed. Figures \ref{fig2}(b) and (c) show, respectively, the typical IPS and APS states observed in experiment.      

\begin{table}[htbp]
\centering
\setlength{\abovecaptionskip}{0pt}%
\setlength{\belowcaptionskip}{10pt}
\caption{By tuning the frequency mismatch, $\Delta f$ (Hz), between the metronomes, the probabilities of generating different states in $100$ realizations. IPS: in-phase synchronization. APS: anti-phase synchronization.}
\begin{tabular}{p{3cm}<{\centering} p{3cm}<{\centering} p{3cm}<{\centering} p{4cm}<{\centering}}
\toprule
$\Delta f$ (Hz) & IPS & APS & Non-synchronization\\
\hline
$3.0\times 10^{-3}$   & 51/100 	& 49/100 & 0\\
$-4.0\times 10^{-3}$   & 26/100   & 74/100 & 0\\
$-14.3\times 10^{-3}$ & 2/100     & 98/100 & 0\\
\hline
\end{tabular}
\label{table1}
\end{table}

We next lift up the clip on metronome 2 by about $1$ cm. This leads to the decreased natural frequency $f_2\approx 1.489$ Hz, and the frequency mismatch is changed to $\Delta f\approx-4.0\times 10^{-3}$ Hz. Having determined $f_2$, we put both metronomes on the board and investigate again the probabilities of generating IPS and APS. As shown in Tab. \ref{table1} (the $2$nd row), the probabilities of generating IPS and APS are $26\%$ and $74\%$, respectively. Lifting up the clip on metronome 2 by about $2$ cm, we have $f_2\approx 1.4787$ Hz and $\Delta f\approx -14.3\times 10^{-3}$ Hz. The probabilities of generating IPS and APS are $2\%$ and $98\%$, respectively, as shown in Tab. \ref{table1} (the $3$rd row). Like the case of $\Delta f\approx 3\times 10^{-3}$ Hz, no non-synchronization state is observed for $\Delta f\approx -4\times 10^{-3}$ Hz and $\Delta f\approx -14.3\times 10^{-3}$ Hz either. Clearly, the probabilities for observing IPS and APS are dependent of the frequency mismatch. More specifically, with the increase of the frequency mismatch, $|\Delta f|$, the probability for generating APS is increased, while is decreased for IPS.  

\section{Theoretical study}

\begin{figure}
\includegraphics[width=0.65\columnwidth]{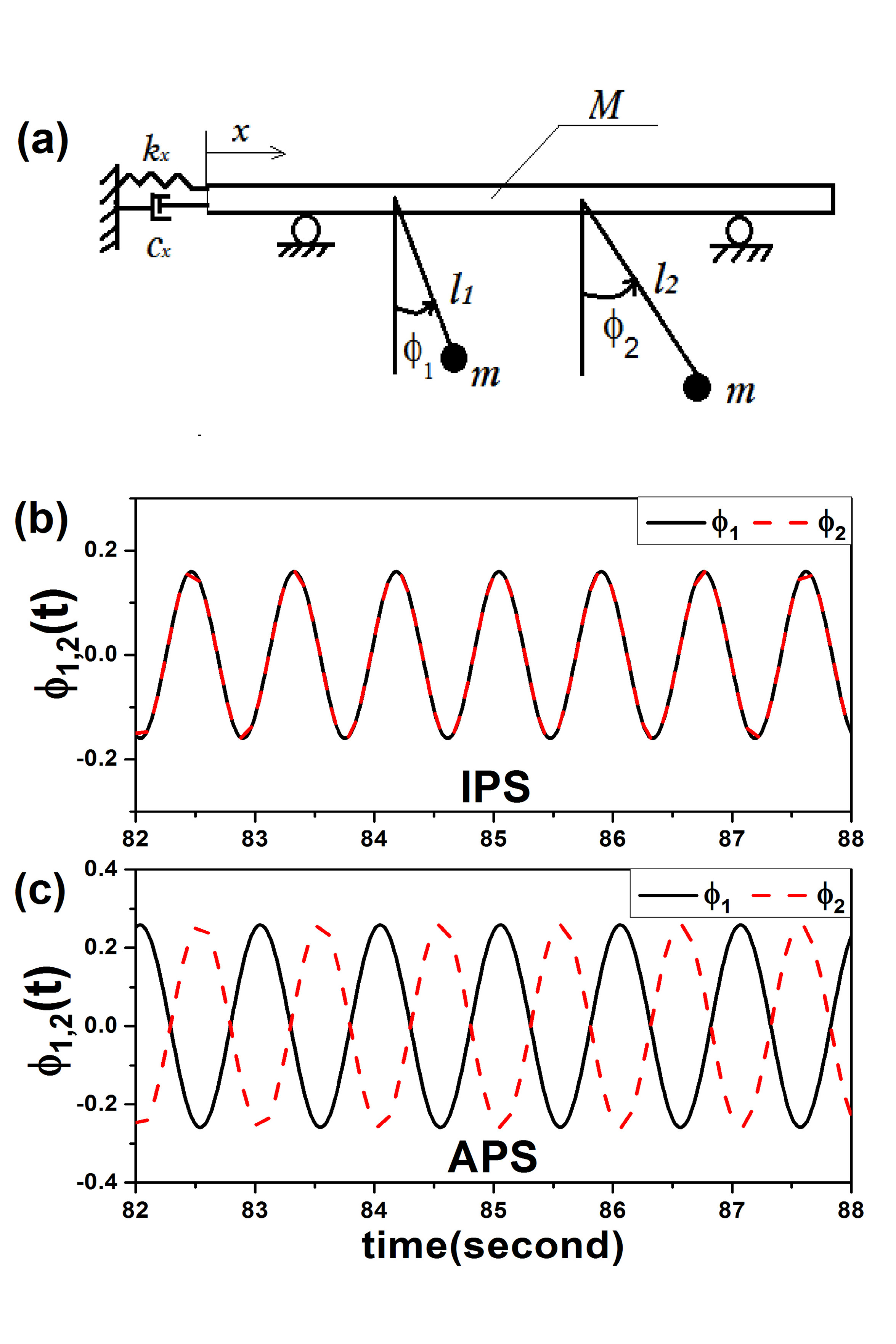}
\caption {(Color online) (a) Sketch of the theoretical model. By identical length of the pendulums, the typical IPS state (b) and APS state (c) observed in numerical simulations.}
\label{fig3}
\end{figure}

How could frequency mismatch facilitate APS while deteriorating IPS? To have a deeper understanding on the impact of frequency mismatch on synchronization, we proceed to investigate the dynamics of coupled metronomes by a theoretical model. Following Refs. \cite{2009 Ulrichs,2012 Wu,2014 Wu,2013 Hu}, we model the experimental setup shown in figure \ref{fig2}(a) by the sketch plotted in figure \ref{fig3}(a). In figure \ref{fig3}(a), the rigid beam and the point pendulums represent, respectively, the plastic board and metronomes in the experimental setup. The mass of the beam is denoted by $M$. $x$ denotes the displacement of the beam from its equilibrium point ($x=0$), $k_{x}$ represents the stiffness coefficient of the spring (so as to keep the beam oscillating around the equilibrium point), and $c_{x}$ denotes the friction coefficient between the beam and the support (the empty cans). The metronomes are of the same mass, $m$, but with different lengths, $l_1\neq l_2$. The instant phases of the pendulums are denoted by $\phi_{1,2}(t)$. To mimic the energy input of the mechanical metronome, we drive each pendulum by the momentum $D_{1,2}$ when $|\phi_{1,2}|$ is smaller to a small angle $\gamma_{N}$. Specifically,  $D_{1,2}=D$ when $0<\phi_{1,2}<\gamma_{N}$ and $\dot{\phi}_{1,2}>0$, $D_{1,2}=-D$ when $-\gamma_{N}<\phi_{1,2}<0$ and $\dot{\phi}_{1,2}<0$, otherwise $D_{1,2}=0$.

The Lagrangian of the theoretical model reads
\begin{equation}\label{e1}
L=\frac{1}{2}[M\dot{x}^{2}+mv_1^2+mv_2^2]+mg(l_{1}\cos\phi_{1}+l_{2}\cos\phi_{2})-\frac{1}{2}k_x x^2,
\end{equation}
with 
\begin{equation}
v^2_{1,2}=[\frac{d}{dt}(x+l_{1,2}\sin\phi_{1,2})]^2+[\frac{d}{dt}(l_{1,2}\sin\phi_{1,2})]^2 \\\nonumber
\end{equation}
and $g$ the gravity constant. Taking into account the friction and driving momentum, we can obtain from the Lagrangian the following equations of the system dynamics:
\begin{eqnarray}\label{e3}
\begin{split}
&ml_{1,2}^{2}\ddot{\phi}_{1,2}+ml_{1,2}\ddot{x}\cos\phi_{1,2}+c_{\phi}\dot{\phi}_{1,2}+mgl_{1,2}\sin\phi_{1,2}=D_{1,2},\\
&(M+2m)\ddot{x}+c_x\dot{x}+k_x x+ml_1(\ddot{\phi}_1\cos{\phi}_1-\dot{\phi}_1^2\sin{\phi}_1)+ml_2(\ddot{\phi}_2\cos{\phi}_2-\dot{\phi}_2^2\sin{\phi}_2)=0.
\end{split}
\end{eqnarray}
Here $c_{\phi}$ represents the damping coefficient of the pendulums. In our theoretical studies, the system parameters are chosen as: $M=10.0$ kg, $m=1.0$ kg, $c_{\phi}=0.01$ N$\cdot$s/m, $c_{x}=5.0$ N$\cdot$s/m, $g=9.81$ m/s$^{2}$, $k_{x}=1.0$ N/m, $D=0.075$ N$\cdot$m, and $\gamma_{N}=\pi/36$. It is noted that this set of parameters are just for the purpose of illustration, which are not derived from the experimental setup directly. As for the lengths of the pendulums, we fix $l_{1}=0.124$ m ($f_1\approx 1.42$ Hz), while changing $l_2$ to adjust the frequency mismatch. (The larger is $l$, the smaller is the natural frequency of the pendulum.) The beam is initially staying at the equilibrium point with $0$ velocity, i.e., $x=0$ m and $\dot{x}=0$ m/s. As in our experiments, the pendulums are released from the static states, and the initial phases of the pendulums are randomly chosen within the range $[-0.5,0.5]$. Eqs. (\ref{e3}) are solved numerically by the fourth-order Runge-Kutta method, with the time step $\delta t=1\times 10^{-3}$ s. After a transient period of $T_{tr}=1\times 10^{3}$ seconds, we compare the phase difference, $\delta \phi=|\phi_1-\phi_2|$, between the pendulums for a period of $T=5\times 10^{2}$ seconds. The criteria for IPS and APS are the same to that of experimental study: the system is regarded as reaching IPS if $\delta \phi<\Delta \phi$ during the period of $T$, reaching APS if $|\delta \phi-\pi|<\Delta \phi$, and reaching non-synchronization otherwise. Still, the detecting threshold for synchronization is chosen as $\Delta \phi=0.3$ rad.    

\begin{figure}
\includegraphics[width=0.85\columnwidth]{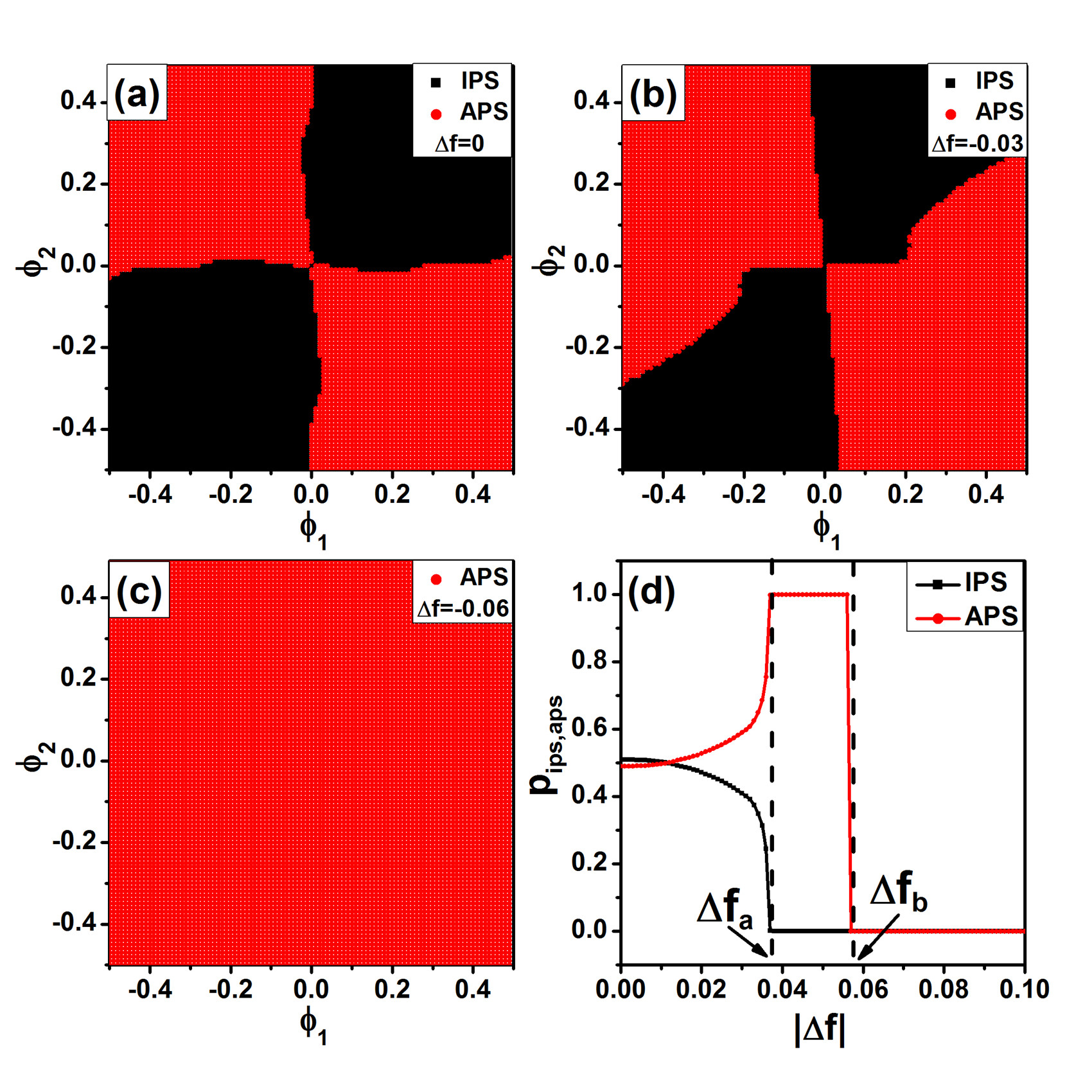}
\caption {(Color online) The impact of frequency mismatch, $\Delta f$, on the attracting basins of IPS (black squares) and APS (red disks). (a) $\Delta f=0$ Hz. (b) $\Delta f=-3.0\times 10^{-2}$ Hz. (c) $\Delta f=-6.0\times 10^{-2}$ Hz. (d) The variations of the normalized basin sizes, $p_{ips}$ and $p_{aps}$, with respect to $|\Delta f|$. $p_{aps}=1$ in the range $\Delta f\in [\Delta f_a,\Delta f_b]$, and is $0$ when $|\Delta f|>\Delta f_b$, with $\Delta f_a\approx 3.7\times 10^{-2}$ Hz and $\Delta f_b\approx 0.58$ Hz.}
\label{fig4}
\end{figure}

We first study the synchronization behavior of two identical pendulums, i.e., setting $l_2=l_1$ in Eqs. (\ref{e3}). To search for all the possible states, we scan over the region $[-0.5,0.5]$ in the two-dimensional phase space $(\phi_1,\phi_2)$ by the resolution $1\times 10^{-2}$ rad, and plot in figure \ref{fig4}(a) the attracting basins of the different states. In figure \ref{fig4}(a), the black squares and red disks denote, respectively, the initial conditions that lead to IPS and APS. Statistically, the larger is the basin of a specific state, the higher is the probability for this state to be generated by an randomly chosen initial condition. Denote $p_{ips}$, $p_{aps}$ and $p_{nsy}$ as the fractions of initial conditions inside the basins of the IPS, APS and non-synchronization states, respectively, we have $p_{ips,aps,nsy} \approx \rho_{ips,aps,nsy}$, with $\rho_{ips,aps,nsy}$ the probability of generating each state. For the numerical results shown in figure \ref{fig4}(a), we have $p_{ips}\approx 0.52$, $p_{aps}\approx 0.48$ and $p_{nsy}\approx 0$. The typical IPS and APS states observed in simulations are shown in figures \ref{fig3}(b) and (c), respectively.
 
We next tuning the natural frequency of the $2$nd pendulum by increasing $l_2$. Increasing $l_2$ to $0.13$ m, we have $f_2\approx 1.39$ Hz and $\Delta f=-3.0\times 10^{-2}$ Hz. The attracting basins of the asymptotic states are plotted in figure \ref{fig4}(b). We have $p_{ips}\approx 0.41$, $p_{aps}\approx 0.59$ and $p_{nsy}\approx 0$. Comparing to the case of identical pendulums [see figure \ref{fig4}(a)], it is seen that the basin of APS is clearly enlarged. Increasing $l_2$ further to $0.134$ m, we have $f_2\approx 1.36$ Hz and $\Delta f=-6.0\times 10^{-2}$ Hz. The basins of the asymptotic states are plotted in figure \ref{fig4}(c). It is seen that all initial conditions in the scanned region lead to APS, i.e., $p_{aps}\approx1$. To investigate systematically the impact of frequency mismatch on the attracting basins, we plot in figure \ref{fig4}(d) the variations of $p_{ips}$ and $p_{aps}$ with respect to $|\Delta f|$. Figure \ref{fig4}(d) shows that as $|\Delta f|$ increases from $0$, the basin of IPS (APS) is gradually decreased (increased), and, at about $\Delta f_a=3.7\times 10^{-2}$ Hz, we have $p_{ips}\approx 0$ and $p_{aps}\approx1$. The value of $p_{aps}$ is staying at $1$ in the range $|\Delta f|\in [\Delta f_a, \Delta f_b]$, with $\Delta f_b\approx 0.58$ Hz. After that, $p_{aps}$ is suddenly decreased from $1$ to $0$, and keeping on $0$ as $|\Delta f|$ increases further. As $p_{ips}=p_{aps}=0$, the pendulums always reach the non-synchronization state for $|\Delta f|>\Delta f_b$. Basin analysis thus indicated that, as observed in the experiments, the probability of generating APS is indeed influenced by the frequency mismatch. 

\begin{figure}
\includegraphics[width=0.65\columnwidth]{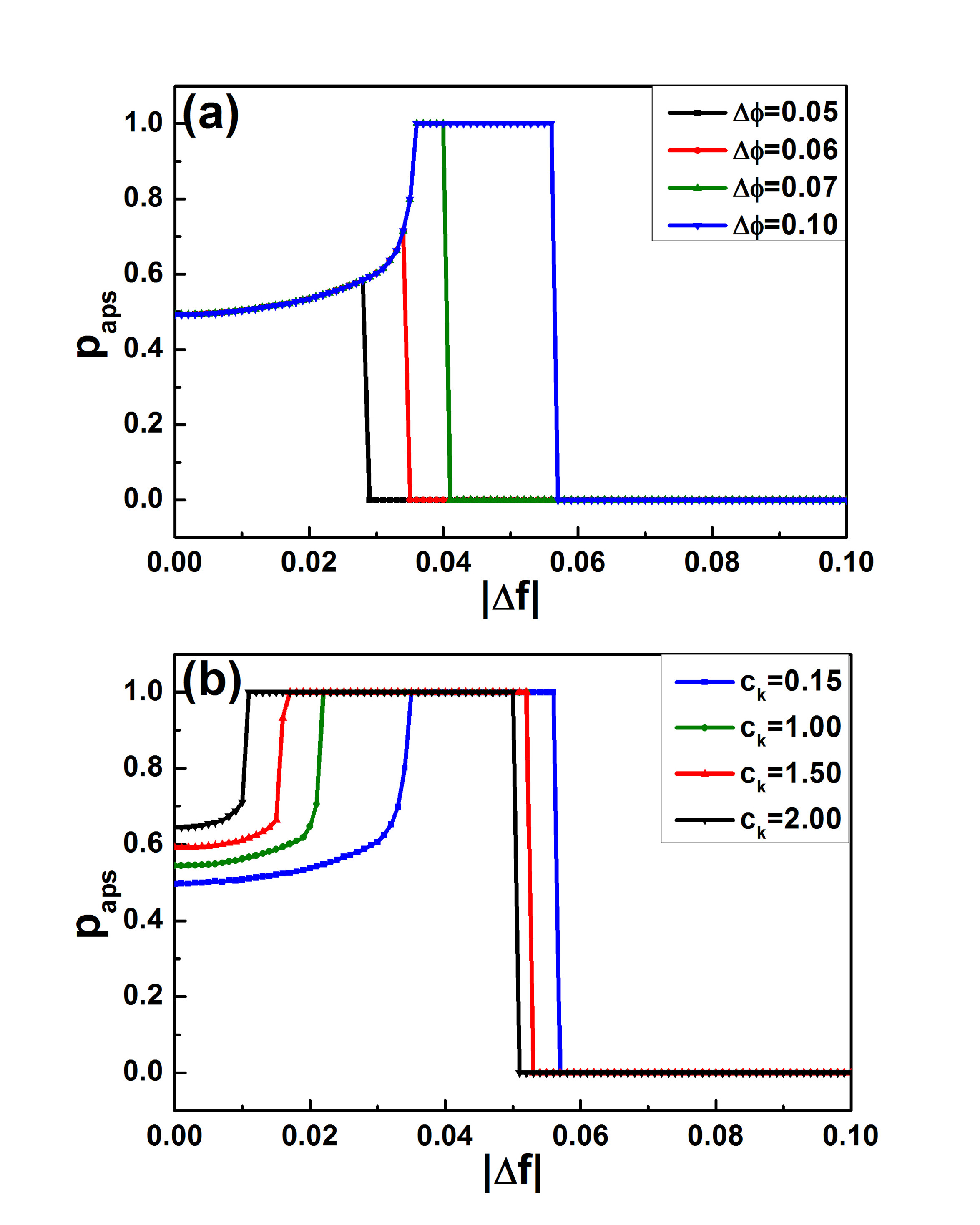}
\caption {(Color online) The impacts of (a) the synchronization criterion, $\Delta \phi$, and (b) the friction coefficient, $c_x$, on the the generating probability of APS.}
\label{fig5}
\end{figure}

We move on to investigate the impacts of the synchronization criterion, $\Delta \phi$, and friction coefficient, $c_x$, on the generation of APS. With the same parameters used in figure \ref{fig4}, we plot in figure \ref{fig5}(a) the variation of $p_{aps}$ with respect to $|\Delta f|$ for different synchronization criteria. It is seen that as $\Delta \phi$ decreases, the desynchronization point, $\Delta f_b$, is moved to the left gradually, whereas before this point $p_{aps}$ is independent of $\Delta \phi$. Numerical results thus suggest that by changing the synchronization criterion, it is only the upper limit of the attracting basin that is affected; before the desynchronization point $|\Delta f_b|$, the basin of APS is always monotonically increased with the frequency mismatch. However, by decreasing $\Delta \phi$, the range over which the basin of APS is enlarged with $|\Delta f|$ is narrowed. (In the extreme case of $\Delta \phi=0$, we have $\Delta f_b=0$, which means that the introduction of any frequency mismatch will result in desynchronization.) Figure \ref{fig5}(b) shows the impact of $c_x$ on the attracting basin of APS. It is seen that by increasing $c_x$, the basin of APS is gradually increased, especially for small $\Delta f$. (Of course, when $c_x$ is too large, the basin of APS will be disappeared. In such a case, neither APS nor IPS will be observed.) The phenomenon depicted in figure \ref{fig5}(b) is in consistent with the finding in Refs.~\cite{2012 Wu,2014 Wu}, where it is shown that the basin of APS is enlarged by increasing $c_x$ slightly. 

\begin{figure}
\begin{center}
\includegraphics[width=0.65\linewidth]{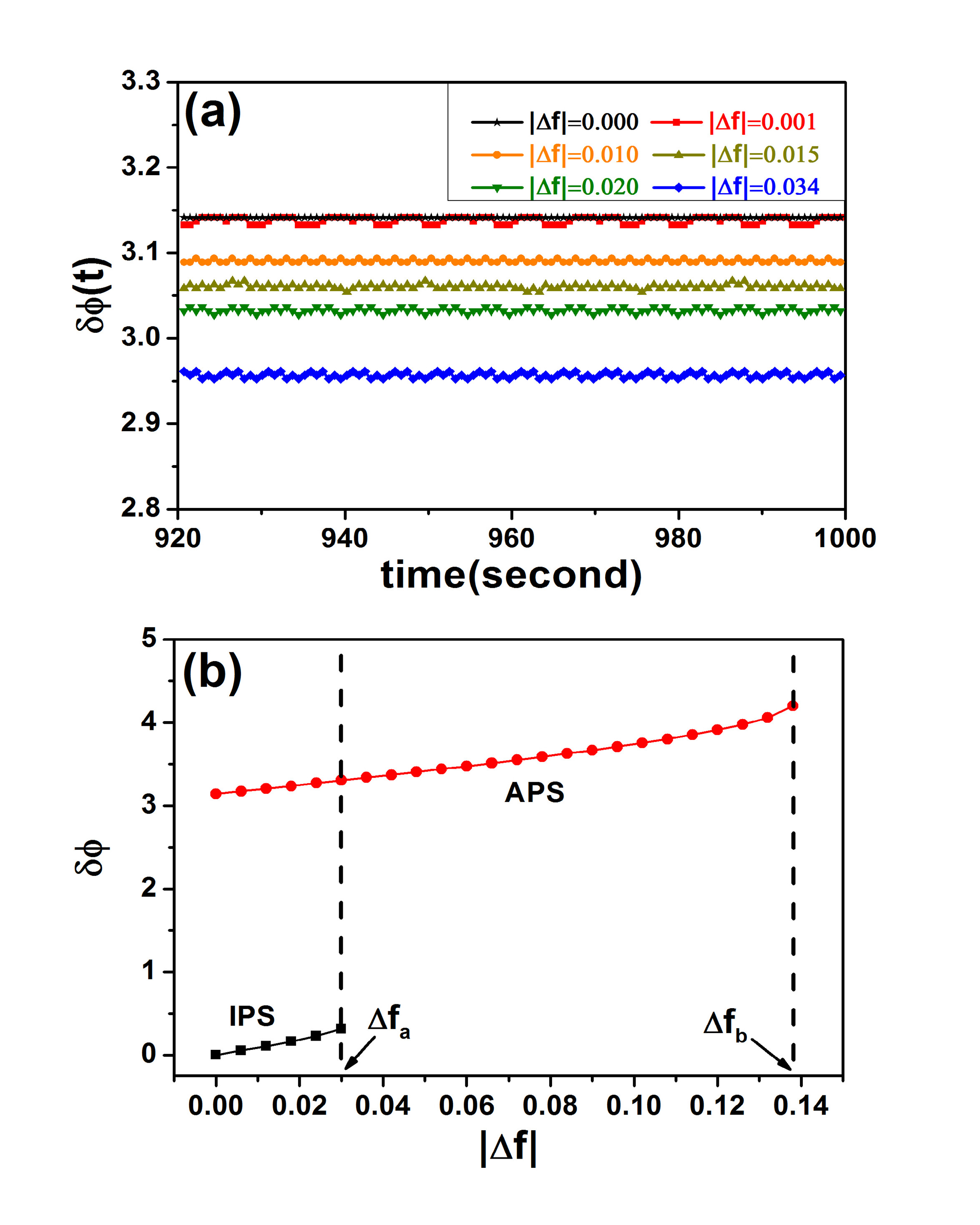}
\caption{(Color online) The trade-off between the frequency mismatch, $|\Delta f|$, and the synchronization precision, $\Delta \phi$. (a) In the APS states, the time evolution of the phase difference, $\delta \phi$, for different $|\Delta f|$. (b) The variations of $\delta \phi$ with respect to $|\Delta f|$ for both IPS and APS states.} \label{fig6}
\end{center}
\end{figure}

Figure \ref{fig5}(a) suggests that the basin of APS is increased at the cost of decreased synchronization precision. To have more details on the trade-off between the generating probability and synchronization precision of APS, we finally investigate the dependence of synchronization precision on frequency mismatch. When $\Delta f=0$ and without noise perturbations, the pendulums can be perfectly synchronized, i.e., $\delta \phi(t)=\phi_1(t)-\phi_2(t)=0$ (IPS) or $\delta \phi(t)=\pi$ (APS) in the asymptotic states. When $\Delta f\neq 0$, the states $\phi_1(t)=\phi_2(t)$ and $\phi_1(t)=\phi_2(t)+\pi$ are no more solutions of the system dynamics. For the latter, the IPS and APS states can only be defined loosely, i.e., $|\delta \phi(t)|<\Delta \phi$ (IPS) or $|\delta \phi(t)-\pi|<\Delta \phi$ (APS), with $\Delta \phi$ the synchronization criterion (threshold or precision). Figure \ref{fig6}(a) shows the time evolution of $\delta \phi$ of the APS state for different values of $\Delta f$. It is seen that after the transient period, $\delta \phi(t)$ is stabilized onto a constant value which, as $|\Delta f|$ increases, is gradually deviated from $\pi$. To have a global picture on the impact of $\Delta f$ on synchronization precision, we plot in figure \ref{fig6}(b) the variation of $\delta \phi$ with respect to $|\Delta f|$ over a large range (the phases are not locked when $|\Delta f|>0.14$). It is seen that for both the IPS and APS states, $\delta \phi$ is monotonically increased with $|\Delta f|$. Figure \ref{fig6}(b) shows also that the introduction of frequency mismatch facilitates APS than IPS, as in IPS the phases are unlocked at $|\Delta f|\approx 0.03$, while in APS the phases are unlocked at $|\Delta f|\approx 0.14$. This observation is in consistent with the results depicted in figure {fig4}(d), where IPS is disappeared at $|\Delta f_a|\approx 3.7\times 10^{-2}$ and only APS is observed within the range $|\Delta f|\in(\Delta f_a,\Delta f_b)$. 

\section{Discussions and conclusion}

In the study of oscillator synchronization, a well-known feature is that the introduction of parameter mismatch will deteriorate synchronization~\cite{SYN:Rev,Kuramoto:Rev,SJ:EPL,AS:EPL}. For example, in coupled phase oscillators described by the Kuramoto model~\cite{Kuramoto:Rev}, the onset of synchronization will be postponed to larger coupling strength as the distribution of the natural frequency is broadened; and in coupled chaotic oscillators, around the critical point the synchronization error is monotonically increased with the increase of parameter mismatch~\cite{SJ:EPL,AS:EPL}. Yet, in realistic situations, parameter mismatch is always unavoidable. Facing this reality, an active topic in synchronization studies has been developing new strategies to reduce the destructing role of parameter mismatch, e.g., by rearranging the oscillators or updating the coupling schemes~\cite{SYNREV:Arenas}. Our present work reveals, from a different viewpoint, the constructive role of parameter mismatch on synchronization. That is, parameter mismatch facilitates APS. This finding provides new perspectives on the collective behavior of coupled oscillators, and might have implications to the functionality and operation of many realistic systems, e.g., the critical role of APS in realizing the working memory of the human brain~\cite{APPL:BRAIN}, the emergence of APS in influenza propagations~\cite{APPL:INF}, and the APS pattern appeared in restoring the oscillation of coupled electrochemical systems from amplitude death~\cite{APPL:ELE}.  (Large scale IPS, on the other hand, implies usually dysfunctions, e.g., epilepsy seizures~\cite{APPL:SEIZE}, and disasters, e.g., bridge collapses~\cite{APPL:BRIDGE}, in realistic situations.)      

The finding that frequency mismatch could promote APS will be helpful to classroom demonstrations. In classroom demonstration of metronome synchronization, it is much difficult to generate APS than IPS. This difficulty makes many instructor chose to demonstrate only IPS in classroom, whereas APS is only briefly mentioned or interpreted according to the experimental photos. In Ref. \cite{2012 Wu}, the authors proposed to enlarge the basin of APS by increasing the friction coefficient, $c_x$, which is realized by tapping paper tissues between the empty cans and the desk. This skill, while is able to increase the generating probability of APS, has the drawbacks of inconvenient operation and non-controllable parameters. In particular, the friction coefficient can not be precisely measured and, more seriously, the friction coefficient is decreased gradually as the cans roll on the tissues. These drawbacks are overcome in our current scheme. First, when the position of the clip is fixed, the natural frequency of the metronome will be determined (time independent), so is the frequency mismatch between the metronomes. Second, the positions of the clips can be adjusted continuously along the pendulum bob, making it possible to freely control the frequency mismatch. Finally, the frequency mismatch can be precisely measured, making it possible to evaluate the impact of frequency mismatch on synchronization in a quantitative manner. We therefore hope that, equipped with the new scheme, the demonstration of APS would not be the ``headache", but the ``showtime" for instructors.  

Summarizing up, we have studied, experimentally and theoretically, the synchronization of two coupled mechanical metronomes, and found that by increasing the frequency mismatch between the metronomes, the probability for generating APS from the random initial conditions can be clearly increased. The study sheds lights on the collective behavior of coupled oscillators, and will be helpful to the classroom demonstration of synchronization.  

Recalling the ``odd sympathy" discovered by Huygens, we have reason to speculate that the pendulums in Huygens' ketch might have been drawn deliberately as different, so as to emphasize the role of frequency mismatch on generating APS. 

\section*{Acknowledgement}
This work was supported by the National Natural Science Foundation of China under the Grant No.~11375109, and by the Fundamental Research Funds for the Central Universities under the Grant No.~GK201601001. 

\appendix*
\section{Method for acquiring the instant phases of metronomes}  

The instant phases of the metronomes, $\phi_{1,2}(t)$, are acquired from the recorded video by software, as follows. First, we treat each frame of the video as a two-dimensional picture, while setting the lower-left corner of the picture as the origin of coordinates. For the camera used in our experiment, the resolution of the picture is $1920\times 1080$ pixels. We then use $(x,y)$, with $x\in [1,1920]$ and $y\in[1,1080]$, to locate the pixels, and assign each pixel with its color value (RGB triplet). For the blue-colored markers sticked on the metronomes, the associated RGB triplet is $00255$. By scanning over the picture, we are able to identify all the blue pixels in the picture. Second, according to the position of the markers, we group the blue pixels into $4$ clusters, with $(x^l_{i},y^l_{i})$ the coordinates of the $i$th pixel in cluster $l$. For the $l$th cluster, we calculate the averaged coordinates $x^l_c=\sum^{n_l}_{i=1}x^l_i/n_l$ and $y^l_c=\sum^{n_l}_{i=1}y^l_i/n_l$ ($n_l$ is the number of pixels in cluster $l$), and define $(x^l_c,y^l_c)$ as the center of this cluster. Third, we pair the centers according to the metronomes, and define the instant phase of the metronome as 
\begin{equation}
\phi=\arctan\frac{y_c-y'_c}{x_c-x'_c},
\end{equation}
with $(x_c,y_c)$ and $(x'_c,y'_c)$ the upper and lower centers associated to the same metronome. Finally, we analyze the video frame by frame by the same process, and obtain the time evolution of the instant phases depicted in Figs. 2(b) and (c). We would like to note that this analysis can be realized by any software, which in our study is MATLAB.

\end{document}